\documentclass[twocolumn,aps,prc,floatfix]{revtex4}
\usepackage{graphicx}
\usepackage{dcolumn}
\usepackage{bm}

\begin{document}


\title{Nucleon momentum gap in asymmetric nuclear matter}

\author{Gao-Chan Yong} 

\affiliation{Institute of Modern Physics, Chinese Academy of Sciences, Lanzhou 730000, China \&\\
School of Nuclear Science and Technology, University of Chinese Academy of Sciences, Beijing 100049, China}

\begin{abstract}

For more than half a century, nucleons are considered to move continuously in the nuclei. Recent electron-scattering experiments indicate about 20$\sim$25\% nucleons in heavier nuclei are involved in the neutron-proton short-range correlations. Nucleons in the nuclei thus have abnormal behavior unlike those in the non-interaction fermi gas. In the neutron-rich nuclei, around the Fermi momentum, the neutron-proton short-range correlations lead to a momentum gap in the proton momentum distribution, which is circumstantially supported by the pionic experimental data. Likewise, there should also be a neutron momentum gap in the proton-rich nuclei. Nucleon momentum gap is thought to have profound and extensive implications in the studies ranging from particle physics to neutron stars as well as ultra-cold atomic gases.

\end{abstract}

\maketitle


It is generally considered that the atomic nucleus, which composes of protons and neutrons, is one of the most complex quantum-mechanical systems in nature.
In an independent-particle model picture, nucleons move in
the average nuclear field created by their mutual attractive interactions with
momenta less than the Fermi-momentum \cite{shell05}.
While recent analysis of electron- and proton-scattering experiments
indicates about 20$\sim$25\% nucleons in heavier nuclei move
with momenta greater than the Fermi-momentum \cite{nature2018,RMP2017,Claudio15,sci14,sci08,pia06,tang03,fomin12,Egiyan06}.

A fraction of nucleons in a heavier nucleus can form pairs with larger relative momenta
and small center-of-mass momenta \cite{pia06,sh07}. This
phenomenon is considered to be caused by the short-range nucleon-nucleon tensor
interactions \cite{tenf05,tenf07}. The nucleon-nucleon short-range
correlations (SRC) in heavier nuclei lead to a high-momentum tail (HMT)
in the single-nucleon momentum distribution \cite{sci14,bethe71,anto88,Rios09,yin13}.
Experimentally, the high-momentum tail of the nucleon momentum distributions appears to decrease as $1/k^{4}$ \cite{henkp415}. And in the HMT of nucleon momentum distribution,
the nucleon components are significantly isospin-dependent. The number of
n-p SRC pairs is about 18 times the p-p or n-n SRC pairs
\cite{sci08,sci14}. In the neutron-rich nuclei, proton thus has a greater
probability than neutron to have momentum greater than the
Fermi-momentum \cite{nature2018}.

Proton transition momentum refers to the starting point of its HMT \cite{yongjump18}. It directly relates to proton's energy distribution in the nuclei.
It is undoubted that in symmetric nuclei, nucleon transition momentum is its Fermi-momentum \cite{sci14}. However, for asymmetric nuclei (in which the number of neutrons is usually larger than that of protons), the transition momentum of minority has not yet been determined.
It is naturally considered that below the Fermi-momentum, for the momentum distribution function n(k), in the nuclear matter, both proton and neutron have constant distributions while above the Fermi-momentum they have $1/k^{4}$ distributions starting from their \emph{respective} Fermi-momenta. However, in the neutron-rich matter the above assumption contradicts the short-range correlation picture of the neutron-proton pairs \cite{sci14}, i.e., the short-range correlated neutron and proton should have the same magnitude of momentum with opposite directions \cite{pia06,sh07}. Extrapolating to an extreme case, such as in neutron-star matter (the ratio of neutron number over proton number may reach 9), neutron and proton would have very unequal Fermi-momenta due to their very different local densities according to the local Thomas-Fermi relation. If the short-range correlated neutron and proton have $1/k^{4}$ distributions starting from their \emph{respective} Fermi-momenta, then they should possess quite different momenta, i.e., proton's momentum would be much smaller than that of neutron. This case not only conflicts with the short-range correlation picture, but evidently contradicts the findings in Ref.~\cite{Rios14}, in which it is shown by the ladder summation techniques in a self-consistent Green's function approach that the integrated single-particle strength for protons is evidently larger than that for neutrons at very high momenta in the very neutron-rich matter.

\begin{figure}[t]
\centering
\includegraphics[width=0.45\textwidth]{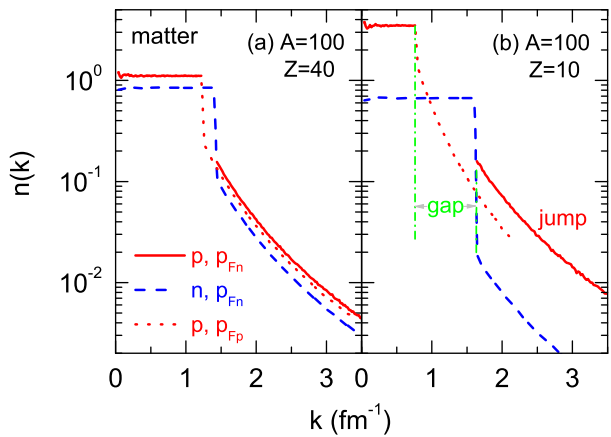}
\caption{ (Color online) Momentum distribution n(k) of nucleon with different starting points of proton $1/k^{4}$ distribution in nearly symmetric matter (A=100 \& Z=40, left panel) and extremely asymmetric matter (A=100 \& Z=10, right panel) with normalization condition $\int_{0}^{k_{max}}n(k)k^{2}dk$ = 1. For neutron, the HMT's starting point is its Fermi-momentum. While for proton's HMT starting point, one case is its Fermi-momentum, the other is the same as neutron's Fermi-momentum labeled by ``jump''. Proton momentum gap refers to the ``jump'' of proton's momentum distribution around its Fermi-momentum.} \label{nk}
\end{figure}
Fig.~\ref{nk} shows proton and neutron momentum distributions in nearly symmetric (panel (a)) and extremely asymmetric matter (panel (b)). It is seen that if one uses respective Fermi-momenta as the HMT's starting points, there is no clear difference in nearly symmetric matter (shown in panel (a)). But the difference becomes prominent in extremely asymmetric matter (shown in panel (b)). If letting proton's starting momentum in the HMT be the same as neutron's, proton's momentum distribution would be noncontinuous. Proton momentum gap refers to the ``jump'' of proton's HMT starting momentum from proton's Fermi-momentum to neutron's Fermi-momentum in the neutron-rich matter. The existence of proton momentum gap guarantees the integrated single-particle strength for protons is evidently larger than that for neutrons at very high momenta in extreme neutron-rich matter \cite{Rios14}.

In the very neutron-rich matter, since the single-particle strength of protons becomes more prominent at high momenta as proton's concentration decreases, the starting momentum of proton's HMT should not be proton's Fermi-momentum. This is because proton's Fermi-momentum would become very small in magnitude as proton's concentration (i.e., its local density) decreases sharply. Theoretically, it is hard to directly obtain the information on the proton momentum gap in momentum distribution in asymmetric matter from microscopic calculations \cite{Rios09,wangp2013,yongjump18}. This is not only because the proton momentum gap is a certain kind of delicate structure of proton momentum distribution around the Fermi momentum, but because around the Fermi momentum the proton momentum distribution has a characteristic of sharp decrease or plunge, which is hard for most microscopic approaches to deal with. As a consequence, the proton momentum gap around its Fermi momentum is most likely to be smoothed out.

Proton momentum gap exists only in the neutron-rich matter, thus should be more clearly shown
in neutron-star matter. For the light-medium nuclei calculated by Wiringa \emph{et al} using the quantum Monte Carlo methods \cite{Wiringa14}, due to very small asymmetry, the so-called proton momentum gap is not clearly demonstrated. For the same reason, one cannot find the proton momentum gap in the calculations for medium to heavy nuclei given by Jan Ryckebusch \emph{et al} using a low-order correlation operator approximation \cite{Ryckebusch15}. The calculations given by A. Rios \emph{et al} using the self-consistent Green's function displayed proton momentum distribution in very neutron-rich matter \cite{Rios14}, but their calculations with finite temperature cannot give fine structure of the nucleon momentum distributions around the Fermi momentum at zero temperature, the proton momentum gap is thus smoothed out. However, the stronger integrated single-particle strength at very high momenta for protons than neutrons in very neutron-rich matter shown in Ref. \cite{Rios14} is a circumstantial evidence of the existence of the proton momentum gap. In this Letter, I make an effort to try to achieve the confirmation of proton momentum gap by using the heavy-ion collisions with the reaction systems in possession of unequal numbers of proton and neutron.


In the present transport model, in colliding nuclei, nucleon's $1/k^{4}$ momentum distribution with a high-momentum tail is adopted. Although the roughly $1/k^{4}$ dependence of high momentum tail of nucleon is in fact observed only for deuteron, assuming the motion of the center of mass of correlated nucleons is isotropic, in heavier nuclei the single nucleon's $1/k^{4}$ dependence of the high momentum tail is averagely used. The short-range correlations cause a rough 20$\sim$25\% depletion of nucleon distribution in the Fermi sea \cite{sci08,sci14}. Specifically, nucleon momentum distribution in the nuclear matter is set to be \cite{yongjump18,henkp415}
\begin{eqnarray}
n(k)=\left\{%
  \begin{array}{ll}
    C_{1}, & \hbox{$k \leq k_{F}$;} \\
    C_{2}/k^{4}, & \hbox{$k_{F} < k < k_{max}$} \\
\end{array}%
\right.
\label{nk}
\end{eqnarray}
with normalization
\begin{equation}\label{norm}
\int_{0}^{k_{max}}n(k)k^{2}dk = 1.
\end{equation}
Where $k_{F}$ is the Fermi-momentum and $k_{max}$= 2$k_{F}$ in this study.
By using the local Thomas-Fermi relation
\begin{equation}
 k_{F_{n,p}}(r)= [3\pi^{2}\rho(r)_{n,p}]^{\frac{1}{3}},
\end{equation}
nucleon momentum distribution in nuclei is then given by \cite{yongli2017}
\begin{equation}
 n_{n,p}(k)= \frac{1}{N,Z}\int _{0}^{r_{max}}d^{3}r\rho_{n,p}(r)\cdot n(k,k_{F}(r)),
\end{equation}
with $N$ and $Z$ being the total numbers of neutron and proton in the nucleus,
$r_{max}$ is the radius of the nucleus and $n(k,k_{F}(r))$ is the nucleon momentum distribution
with local Fermi-momentum $k_{F}(r)$.
With the above settings, the nucleon momentum initialization of $^{56}$Fe fits experimental data quite well \cite{yongli2017}.
The n-p dominance in the short-range correlations causes protons have larger probability than neutrons with momenta greater than the Fermi momentum in the neutron-rich matter \cite{nature2018,sargnp14}.
The n-p dominance model demands the number of neutrons is equal to that of protons in the HMT (the n-p dominance model neglects 2\% n-n and p-p SRCs which weaken the proton momentum gap very slightly). This is achieved via
\begin{equation}\label{frac20}
\int_{k_{F}}^{k_{max}}n^{HMT}(k)k^{2}dk \bigg/ \int_{0}^{k_{max}}n(k)k^{2}dk = 20-25\% (1\pm \delta),
\end{equation}
where ``+'' for protons and ``-'' for neutrons, $\delta$ = $(\rho_{n}-\rho_{p})/\rho$ is the asymmetry.

Pion production in nucleus-nucleus collisions at intermediate energies is modeled in the framework of the isospin-dependent Boltzmann-Uehling-Uhlenbeck transport model \cite{bali1991,npa2002,pion2017}. In this model, pion is produced via the isospin-dependent nucleon-nucleon elastic and inelastic scatterings. Nucleonic resonance may be produced in nucleon-nucleon scatterings. Once produced, the resonance would take part in collision process with all other baryons. Meanwhile, the resonance may decay into nucleon and pion, and the pion may also collide with nucleon and further form resonance. At the last stage of the reaction, all the residual resonances decay into nucleons and pions. More details on the pion production in the model can be found in the literature \cite{pion2017}. In this model, the isospin and density plus momentum-dependent single particle potential and the in-medium cross sections for the baryon-baryon scatterings are used. The isospin dependent both initialization and fermion pauli-blockings are also taken into account. Particularly, effects of the nucleon-nucleon short-range correlations are involved into this transport model \cite{yongp4,yongprc2016,yongli2017}. The present model has been successfully used in many studies relating to pion production in heavy-ion collisions at intermediate energies.


\begin{figure}[t]
\centering
\includegraphics[width=0.45\textwidth]{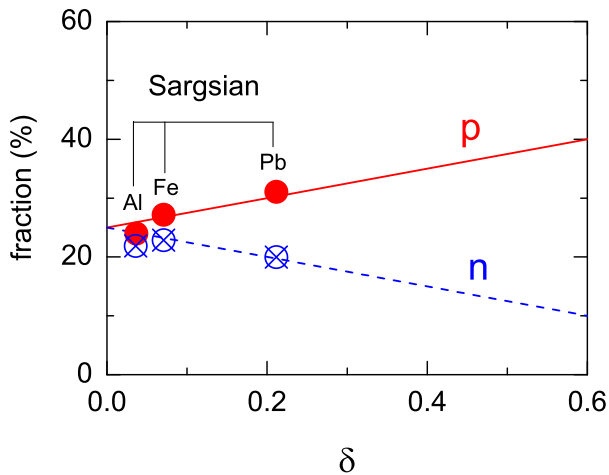}
\caption{ (Color online) Fractions of short-range correlated neutrons and protons as a function of asymmetry. Symbols denote $Al$, $Fe$ and $Pb$ nuclei analyses (solid symbols are for protons and hollows for neutrons) of Sargsian \cite{sargnp14}.} \label{frac}
\end{figure}
Fig.~\ref{frac} demonstrates the fractions of neutrons and protons as a function of asymmetry $\delta$ ($\delta= \frac{\rho_{n}-\rho_{p}}{\rho_{n}+\rho_{p}}$, $\rho_{n}$ and $\rho_{p}$ respectively denote neutron and proton densities). Also shown are the high momentum nucleon fractions in several typical finite nuclei based on the analyses of Sargsian \cite{sargnp14}. From Fig.~\ref{frac}, it is clearly seen that as increase of the asymmetry, the fraction of protons in the HMT increases gradually while the fraction of correlated neutrons decreases. The nucleon fractions in the HMT shown here are in fact the integral results of the nucleon momentum distributions demonstrated in Fig.~\ref{nk}. The ``jump'' of proton momentum results in the change of the proton kinetic energy distribution in neutron-rich nuclei, thus definitely causes dynamical effects in heavy-ion collisions at lower beam energies in which nucleon fermi motion plays an important role.

\begin{figure}[thb]
\centering
\includegraphics[width=0.45\textwidth]{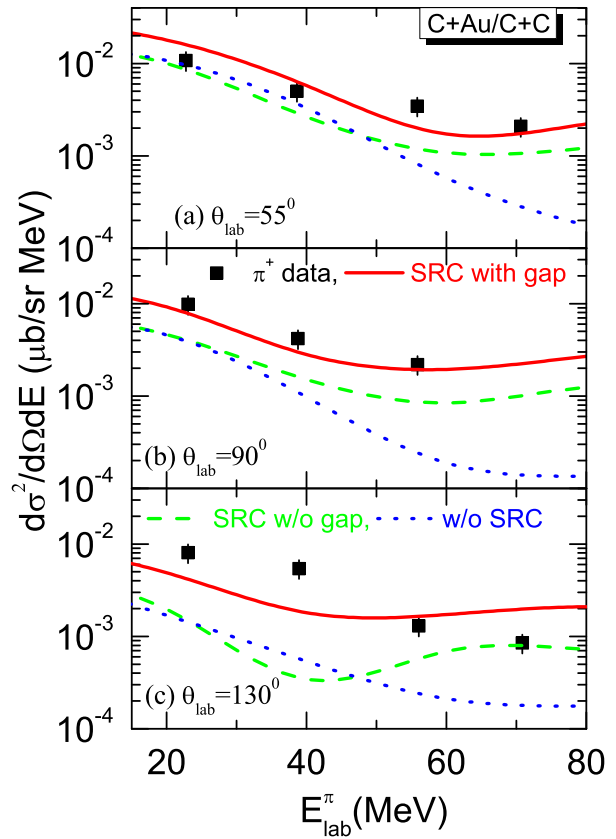}
\caption{ (Color online) Doubly differential cross sections of $\pi^{+}$ production as a function of pion kinetic energy in the C+Au collisions at 85A MeV versus the total cross section of $\pi^{0}$ production in the C+C collisions at 84A MeV. The solid, dash and dot lines show the transport model calculations by using the nucleon short-range correlations with (solid) and without (dash) proton momentum gap and the case without short-range correlations, respectively. Date are taken from Refs. \cite{cern1982,noll1984}.} \label{datacom}
\end{figure}
To confirm proton momentum gap in the neutron-rich nuclei by using heavy-ion collisions, it is preferable to make far below the sub-threshold $\pi^{+}$ production (the threshold energy of pion production is 290 MeV) measurements. This is because the far below sub-threshold $\pi^{+}$ production is closely related to the Fermi motion of proton in nuclei. $\pi^{+}$ production is mainly from proton-proton collision \cite{Sto86}, and the Fermi energy of nucleon plays an important role for far below the sub-threshold $\pi^{+}$ production. $\pi^{+}$ production in heavy-ion collisions is therefore an ideal probe of the proton momentum gap in the neutron-rich nuclei. The measurements of the far below sub-threshold pion production in heavy-ion collisions in fact had been studied about 35 years ago. At that time, the energies of the colliding nuclei provided by accelerators were generally not high and the production mechanism of pion was in debate \cite{cern1982,nag1982,noll1984,bra1984}. To study the proton momentum gap, calculations of the doubly differential cross sections of $\pi^{+}$ as a function of pion kinetic energy in the C+Au collisions at 85A MeV are carried out using my transport model by switching on and off the proton momentum gap as well as the nucleon-nucleon short-range correlations. To reduce systematic errors, such as theoretical uncertainties of the in-medium inelastic cross section and experimental measurement errors, ratios of the pion productions in the two reaction systems, i.e., $\pi^{+}$ production in the C+Au collisions versus $\pi^{0}$ production in the C+C collisions are analyzed. The results are then compared with the available pion data ratio, which are respectively taken from Ref.~\cite{cern1982} and Ref.~\cite{noll1984}. From Fig.~\ref{datacom}, it is seen that there are large differences between my theoretical results and the experimental data without considering SRC in the transport model. By adding SRC in the model, however, my results approach but still cannot well reproduce the data. Only with the proton momentum gap in the model can one obtain good agreement between theory and experiments. Fig.~\ref{datacom} indicates the existence of the proton momentum gap in the neutron-rich nuclei.

\begin{figure}[t]
\centering
\includegraphics[width=0.45\textwidth]{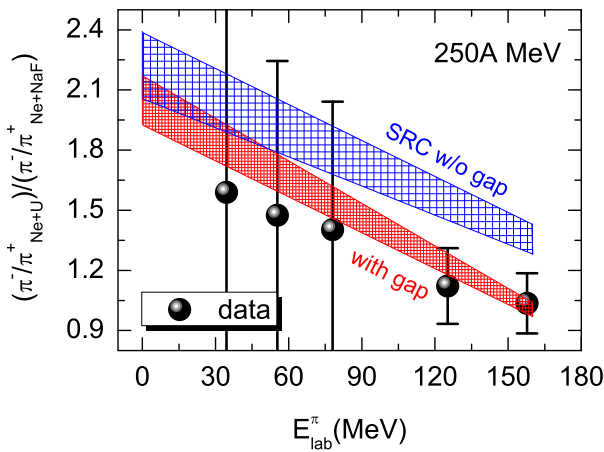}
\caption{ (Color online) Double ratios of the $\pi^{-}/\pi^{+}$ yields vs pion kinetic energy in the reactions of Ne+U and Ne+NaF at the beam energy of 250A MeV with and without the proton momentum gap. Data are taken from Ref. \cite{wben79}.} \label{rpionj}
\end{figure}
To further reduce the systematic errors from the used transport model and experiments, the double ratios of the yields of $\pi^{-}$ and $\pi^{+}$ as a function pion kinetic energy from the two reactions of Ne+U and Ne+NaF at 250A MeV with and without the proton momentum gap are calculated and compared with the experimental data in Ref. \cite{wben79}. Since the $\pi^{-}/\pi^{+}$ yields ratio is sensitive to the symmetry energy, in the calculations, a mildly soft symmetry energy with a slope of about 58.7 MeV at saturation density is employed, which is the most probable form from 55 analyses \cite{Oertel17}. From Fig.~\ref{rpionj}, it is seen that the double ratios of the $\pi^{-}/\pi^{+}$ yields overall decrease as the increase of pion kinetic energy and the splitting of the double ratios becomes prominent at high kinetic energies with and without the proton momentum gap. This is understandable since the proton momentum gap in neutron-rich matter mainly affects the high-energy nucleons thus the energetic pions. Fig.~\ref{rpionj} demonstrates the experimental data supports the existence of the proton momentum gap in the neutron-rich matter.


To summarize, due to the neutron-proton short-range correlations, in the neutron-rich nuclei there in principle should exist proton momentum gap in nucleon momentum distribution. Circumstantial evidence on the existence of the proton momentum gap in the neutron-rich nuclei is provided in the production of $\pi^{+}$ and $\pi^{-}/\pi^{+}$ ratio in the neutron-rich C+Au and Ne+U collisions at low incident beam energies. To reduce the systematic errors, neutron-deficient systems are used as comparisons.
Likewise, there should also exist neutron momentum gap in the proton-rich nuclei or matter. The nucleon momentum gap may change our understanding on the nucleon motion in the nuclei and thus has fundamental implications in the research fields relevant to nucleon motion.

The studies of the neutron-proton short-range interactions and the proton momentum gap could have broad implications on the studies of, such as, the nuclear few-body and many-body systems, the cooling rate and the equation of state of neutron stars, the double-beta decay rate of nuclei, the neutrino-nucleus reactions studying the nature of the electro-weak interaction, the modifications of the nuclear parton distribution functions and the EMC effect as well as the imbalanced ultra-cold atomic gases \cite{nature2018,RMP2017,sci14,sci08}.


This work is supported in part by the National Natural Science Foundation of China under Grant No. 11775275 and the Strategic Priority Research Program of Chinese Academy of Sciences (No. XDB34000000).

\end{document}